# Prediction of electrically-induced magnetic reconstruction at the manganite/ferroelectric interface

J. D. Burton[†] and E. Y. Tsymbal[‡]

*Department of Physics and Astronomy, Nebraska Center for Materials and Nanoscience, University of Nebraska, Lincoln, Nebraska 68588-0111, USA*

The control of magnetization via the application of an electric field, known as magnetoelectric coupling, is among the most fascinating and active research areas today. In addition to fundamental scientific interest, magnetoelectric effects may lead to new device concepts for data storage and processing. There are several known mechanisms for magnetoelectric coupling that include intrinsic effects in single-phase materials, strain induced coupling in two-phase composites, and electronically-driven effects at interfaces. Here we explore a different type of magnetoelectric effect at a ferromagnetic-ferroelectric interface: magnetic reconstruction induced by switching of electric polarization. We demonstrate this effect using first-principles calculations of a La$_{1-x}$A$_x$MnO$_3$/BaTiO$_3$ (001) interface, where $A$ is a divalent cation. By choosing the doping level $x$ to be near a transition between magnetic phases we show that the reversal of the ferroelectric polarization of BaTiO$_3$ leads to a change in the magnetic order at the interface from ferromagnetic to antiferromagnetic. This predicted electrically-induced magnetic reconstruction at the interface represents a substantial interfacial magnetoelectric effect.

## I. INTRODUCTION

Magnetoelectric (ME) materials have recently attracted significant interest due to the possibility of controlling magnetic properties by electric fields.[1,2] In a broad definition, ME phenomena include not only the cross coupling between magnetic and electric order parameters,[3] but also involve related effects such as electrically-controlled magneto-crystalline anisotropy,[4-7] exchange bias,[8,9] and spin transport.[10-14] Tailoring these phenomena by electric fields opens exciting avenues for the design of new data storage and processing devices.

There are several mechanisms giving rise to magnetoelectric effects (for recent reviews see Ref [15] and Ref [16]). An intrinsic magnetoelectric coupling occurs in compounds with no time-reversal and no space-inversion symmetries.[17] In such materials, an external electric field displaces the magnetic ions, eventually changing the exchange interactions between them and hence the magnetic properties of the compound.[18,19] A different mechanism of magnetoelectric coupling may occur in composites of piezoelectric (ferroelectric) and magnetostrictive (ferro- or ferrimagnetic) compounds. In such structures an applied electric field induces strain in the piezoelectric constituent which is mechanically transferred to the magnetostrictive constituent, where it induces a magnetization.[20-23] The importance of composite multiferroics follows from the fact that none of the existing single phase multiferroic materials combine large and robust electric and magnetic polarizations at room temperature.[24]

At ferromagnet/insulator interfaces the ME effect may originate from purely electronic mechanisms. It was predicted that displacements of atoms at the ferromagnet/ferroelectric interface caused by ferroelectric instability alter the overlap between atomic orbitals at the interface which affects the interface magnetization.[25] This produces a ME effect that manifests itself in the abrupt change of the interface magnetization caused by ferroelectric switching under the influence of applied electric field. The ME effect due to the interface bonding mechanism is expected to play a role for Fe/BaTiO$_3$ [25,26], Co$_2$MnSi/BaTiO$_3$ [27], and Fe$_3$O$_4$/BaTiO$_3$ [28] interfaces. Another electronic mechanism for an interface ME effect originates from spin-dependent screening.[29,30] In this case, an applied electric field produces an accumulation of spin-polarized electrons or holes at the metal/insulator interface resulting in a change of the interface magnetization.[31,32] This mechanism is also relevant to ferromagnet/ferroelectric interfaces where screening of the polarization charge alters with ferroelectric polarization orientation producing a change in surface magnetization, as was recently predicted for the SrRuO$_3$/BaTiO$_3$ interface.[33] The experimental indication of the carrier-induced ME coupling was recently found for the PbZr$_{0.2}$Ti$_{0.8}$O$_3$ (PZT)/La$_{0.8}$Sr$_{0.2}$MnO$_3$ interface.[34]

In this work we explore a different type of magnetoelectric effect at a ferromagnet/ferroelectric interface: magnetic reconstruction induced by switching of electric polarization. The effect is fundamentally different since the change in surface magnetization arises not from a variation in the *magnitude* of local magnetic moments, but instead due to a change in how these magnetic moments are ordered near the interface. It is well known that the doped La-manganites, La$_{1-x}$A$_x$MnO$_3$, ($A$ = Ca, Sr, or Ba) posses a rich phase diagram as a function of hole concentration ($x$) and temperature that includes metal-insulator transitions as well as the colossal magneto-resistance effect.[35] Of particular interest for the phenomenon predicted here is the fact that the La-manganites go through a series of magnetic phases, from antiferromagnetic to ferromagnetic and back to antiferromagnetic as $x$ is varied from 0 to 1. These transitions are generally attributed to the competition between the super exchange interaction, which favors antiparallel alignment of neighboring Mn magnetic moments, and the double exchange interaction, which favors parallel alignment.[36] Therefore changing the population of electrons that mediate the double exchange leads to the rich series of magnetic phase transitions.

In addition to explicit chemical doping, carrier concentration can also be modulated electrostatically, opening the possibility to dramatically alter the properties of the manganite in a field effect device.[37] By choosing $x$ to reside near a magnetic phase transition it should then be possible to drive the magnetic order back and forth between different phases through electrostatic screening. Here we demonstrate this effect from first-principles for a La$_{1-x}$A$_x$MnO$_3$(LAMO)/BaTiO$_3$ interface, where the screening charge density in the manganite near the interface depends on the orientation of the ferroelectric polarization in the adjacent BaTiO$_3$ (BTO) layer. Such a phenomenon constitutes a substantial and robust ME effect that is of considerable interest to the area of electrically controlled magnetism.

## II. STRUCTURE AND METHODS

We explore this ME effect at the manganite/ferroelectric interface using first-principles calculations based on density-functional theory (DFT). Fig. 1 shows the supercell which is used in the calculations. It consists of 5.5 unit cells of LAMO[38] and 4.5 unit cells of BTO stacked along the [001] direction of the conventional perovskite cell, assuming the typical $A$O-$B$O$_2$ stacking sequence of perovskite heterostructures. We treat the La-$A$ substitutional doping in a generic way, known as the virtual crystal approximation,[39] by considering this site to be occupied by a fictitious atom with atomic number $57x + 56(1 - x)$. Technically this is closest to representing the Ba-doped manganite. However, since the primary consideration is the change in valence, this virtual crystal is also a reasonable approximation for describing Sr or Ca substitution. Unless stated otherwise, all results presented here assume the nominal manganite concentration $x = 0.5$, chosen to be near the FM-AFM phase transition,[40] where a switchable magnetic reconstruction occurs. The BTO layer is terminated on both ends with BaO atomic layers. The polarization of BTO leads to two distinct interfaces in the supercell: one where polarization is pointing into the interface and one where the polarization points away from the interface (see Fig. 1). Comparing the properties of these two interfaces therefore yields information about the changes at a single interface induced by a reversal of the polarization.

DFT calculations are performed using the plane-wave pseudopotential method implemented in Quantum-ESPRESSO.[41] The exchange-correlation is treated using the generalized gradient approximation (GGA).[42] All calculations use an energy cutoff of 400 eV for the plane wave expansion. Atomic relaxations are performed using a 6×6×1 Monkhorst-Pack grid for k-point sampling and atomic positions are converged until the Hellmann-Feynman forces on each atom became less than 20 meV/Å. The in-plane lattice constant of the supercell is constrained to the calculated (GGA) value for bulk cubic SrTiO$_3$, $a = 3.937$Å, to simulate epitaxial growth on a SrTiO$_3$ substrate. Relaxations for the supercell are carried out by performing separate atomic relaxation calculations of bulk LAMO and BTO with this in-plane constraint and then using the corresponding tetragonal structures to construct the supercell, which is then fully relaxed. Subsequent total energy calculations are performed using either a 20×20×2 grid or a 14×14×2 grid for those configurations requiring a √2×√2 in-plane expansion to accommodate the magnetic order. The pseudopotentials for the fictitious La$_{1-x}$A$_x$ atoms are created using Vanderbilt's Ultra-Soft-PseudoPotential generation code.[43,44]

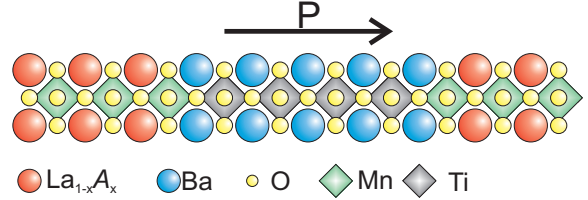

FIG. 1. Atomic structure of the (La$_{1-x}$A$_x$MnO$_3$)$_{5.5}$|(BaTiO$_3$)$_{4.5}$ supercell used in the calculations. The arrow indicates the direction of the polarization, $P$, in the BTO.

Calculations for constrained bulk LAMO find a tetragonal distortion of $c/a = 0.9931$ ($x = 0.5$) and $c/a = 1.0046$ ($x = 0.33$).[45] This small tetragonal distortion is consistent with the small lattice mismatch < 1% between SrTiO$_3$ and La$_{1-x}$Ba$_x$MnO$_3$ found in experiment.[46] Calculations for constrained BTO result in $c/a = 1.057$. The in-plane constraint ensures that the ferroelectric polarization of the BTO is oriented perpendicular to the plane. This is in contrast to previous calculations for LaMnO$_3$/BaTiO$_3$ superlattices where the in-plane lattice constant was not constrained and the ferroelectric polarization in the BTO was found to lie in the plane.[47] Using the Berry phase method we obtain $P = 50\mu C/cm^2$ for the polarization of the bulk BTO, which is well within the range achievable experimentally through strain modulation.[48]

Fig. 2 shows the layer-resolved polar-displacements near the interface. These are quantified by the relative shift of the metal (cation) and the oxygen (anion) atoms in each atomic layer parallel to the interface. The displacements in the BTO are the typical soft-mode distortion that gives rise to ferroelectric polarization. There are also substantial polar-distortions near the interface in the LAMO, and the sign and magnitude of these polar-distortions depend on the polarization orientation, in essence constituting an ionic contribution to the screening of the bound ferroelectric polarization charges at the interface. These distortions are reminiscent of what was predicted for a free $La_{0.7}Sr_{0.3}MnO_3$ surface.[49]

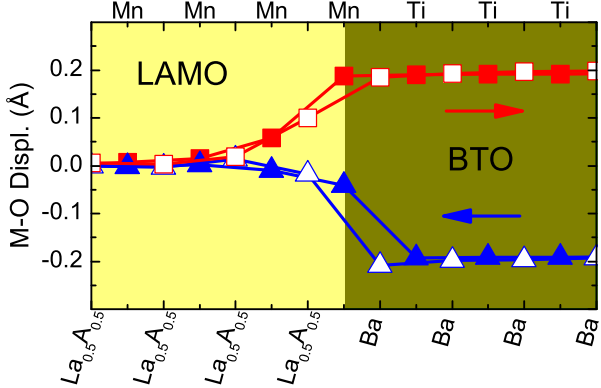

FIG. 2. Metal-Oxygen (M-O) displacements near the LAMO/BTO interface for the states with polarization pointing away from (triangles) and toward (squares) the interface. The solid points are $BO_2$ displacements ($B$ = Mn or Ti) and the open points are A-O displacements (A = $La_{0.5}A_{0.5}$ or Ba).

## III. MAGNETIC RECONSTRUCTION

We examine the magnetic ordering at the LAMO/BTO interface by performing self-consistent supercell calculations for different interfacial magnetic configurations and comparing their energies to the FM state. In spite of the general view that the La-manganites are "strongly-correlated", we expect our DFT calculations to be well suited for determining the magnetic order for doping around $x = 0.5$. In fact it was recently demonstrated that including an additional Hubbard $U$ parameter, which is the most common route to correct DFT calculations of strongly-correlated systems, actually impairs agreement with experiment for the transition from FM to A-type AFM ordering around $x = 0.5$.[50] Indeed, several groups in recent years have applied DFT calculations to the study of magnetic interactions in the doped La-manganites where they find quite reasonable agreement with known experimental results.[50-53] In addition we found that the Fermi level lies well within the band gap of the BTO, which can sometimes be an issue in DFT calculations of metal/insulator interface systems.

We consider interfacial configurations incorporating several stable AFM orderings known for manganites, such as A-type, C-type or G-type.[54] A-type order consists of FM ordered (001) planes of Mn moments that align antiparallel with neighboring planes along the [001] direction. C-type order has AFM alignment between Mn nearest neighbors in the (001) plane, but FM alignment between neighbors along the [001] direction. G-type order has AFM alignment between all six nearest Mn neighbors. (See Fig. 1(a) of Ref. [54])

The resulting magnetic configurations, energies, and magnetic moments are given in Table I. The most striking result is the different magnetic order at the interface favored for opposite polarization orientations of BTO. For polarization pointing into the interface, the FM configuration has minimum energy. For polarization pointing away from the interface, however, there is a transition to the $A_2$ configuration, corresponding to two unit-cells of A-type AFM order at the interface, as shown schematically in Fig. 3. This leads to a net change in interface magnetization of $\Delta M = 7.05\mu_B/a^2$. Notice however that the $A_3$ configuration yields an increase in energy compared to $A_2$, indicating that this effect is limited to the first two unit-cells of LAMO.

TABLE I. Magnetic configuration at the LAMO/BTO interface. The vertical line, |, indicates the interface; the + indicates a spin-up Mn site; the − indicates a spin-down Mn site; and the arrows → (←) indicate polarization orientation. $\Delta E$ is the energy with respect to FM ordering throughout the supercell. $m$ denotes magnetic moments on the Mn sites of the three atomic monolayers nearest the interface.

| Magn./FE config. | | $\Delta E$ (meV) | $m$ ($\mu_B$) |
|---|---|---|---|
| FM | + + + + + + \| ← | 0 | 3.33, 3.42, 3.41 |
| $A_1$ | + + + + + − \| ← | 97 | 3.32, 3.42, −3.40 |
| C | + + + + + + \| ← | 210 | 3.33, 3.41, 3.42 |
|   | + + + + + − \| ← |     | 3.32, 3.42, −3.39 |
| FM | + + + + + + \| → | 0 | 3.25, 3.12, 2.96 |
| $A_1$ | + + + + + − \| → | −16 | 3.25, 3.11, −2.94 |
| $A_2$ | + + + + − + \| → | −29 | 3.23, −3.07, 2.95 |
| $A_3$ | + + + − + − \| → | −21 | −3.16, 3.10, −2.95 |
| C | + + + + + + \| → | 45 | 3.25, 3.11, 2.97 |
|   | + + + + + − \| → |    | 3.25, 3.12, −2.94 |

We find that for both polarization orientations ferromagnetic order within the (001) planes remains stable. As follows from Table I, assuming C-type antiferromagnetic ordering, i.e. a checkerboard arrangement of Mn spins, on the first unit-cell increases the energy for both orientations of the polarization. This stability of the in-plane magnetic ordering precludes the possibility of further C-type or G-type magnetic order.

The change in net surface magnetization due to the transition from FM order to the $A_2$ state constitutes a large interfacial ME effect. Similar to previous work[30] we estimate a figure of merit for the effect known as the surface magnetoelectric coefficient, $\alpha_s$, which is defined by $\mu_0 \Delta M = \alpha_s E$, where $E$ is an applied electric field. Although the effect predicted here is non-linear due to the hysteretic dependence of the polarization on electric field, we can still use $\alpha_s$ as a figure of merit for the ME coupling. Assuming that the applied electric field is given by the coercive field of a BaTiO$_3$ thin film, i.e. $E \sim 10^5$ V/cm, we find $\alpha_s$ = 5.3×10$^{-9}$ Gcm$^2$/V. For comparison the ME coefficient at Fe/BaTiO$_3$ interfaces is found to be ~2×10$^{-10}$ Gcm$^2$/V, using the same coercive field.[25] Similar coefficients in the range of $\alpha_s$ = 0.7-3.0×10$^{-10}$ Gcm$^2$/V were predicted at the Co$_2$MnSi/BaTiO$_3$ [27], Fe$_3$O$_4$/BaTiO$_3$ [28] and SrRuO$_3$/BaTiO$_3$ [33] interfaces depending on the details of the interface termination. Surface ME coefficients in non-ferroelectric systems are predicted to be significantly smaller: ~2×10$^{-14}$ Gcm$^2$/V at the surface of the elemental 3$d$ ferromagnets,[30] a universal constant of 6.44×10$^{-14}$ Gcm$^2$/V at the surface of a half-metal[32] and ~2×10$^{-12}$ Gcm$^2$/V at the SrRuO$_3$/SrTiO$_3$ interface.[31] An effect comparable to the one we propose here is predicted to occur in an epitaxial thin film of insulating EuTiO$_3$, where an electric field is expected to induce an AFM to FM transition, although due to an entirely different mechanism.[19]

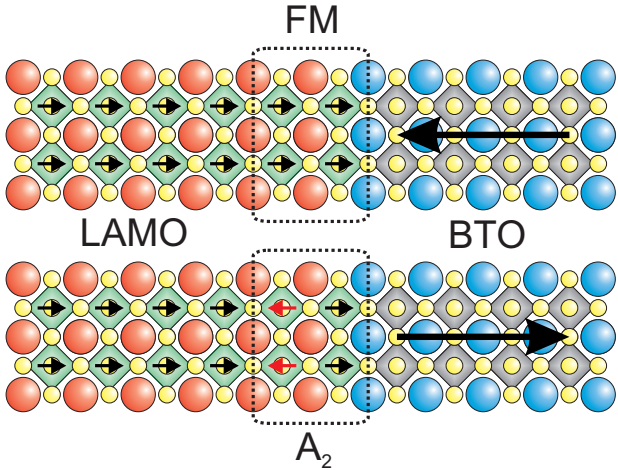

FIG. 3. Electrically-induced magnetic reconstruction at the LAMO/BTO interface: the predicted change in order of the Mn magnetic moments (small arrows) from ferromagnetic (FM) to A-type antiferromagnetic (A$_2$) as the ferroelectric polarization (large arrows) of the BTO is reversed.

## IV. DISCUSSION

The origin of the interfacial magnetic reconstruction is revealed by examining the effects of polarization reversal on the electronic structure at the interface. The polarization in the BTO produces bound charges at the interface which are, in turn, screened by the metallic LAMO. Depending on polarization orientation, this leads either to depletion or accumulation of electrons at the interface. This is evident from Fig. 4 where we plot the local density of states (LDOS) on the interfacial LAMO unit cell assuming FM ordering. For polarization pointing away from the interface, there is a negative bound charge which is screened by reducing the electron population, leading to an upward shift of the LDOS. The opposite occurs when the polarization is pointing into the interface, leading to a downward shift of the LDOS.

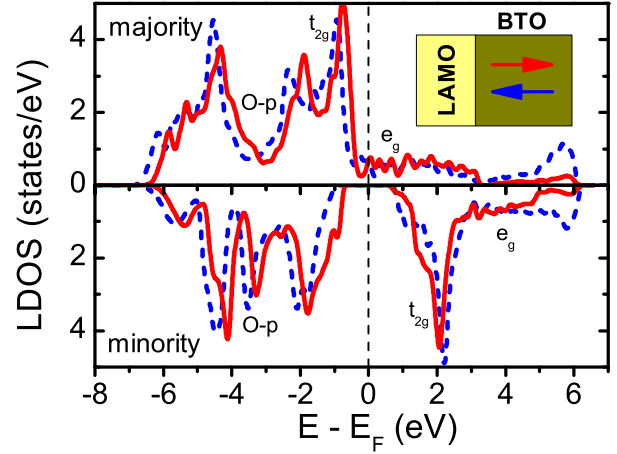

FIG. 4. Spin-resolved local density of states (LDOS) of the interfacial LAMO ($x$ = 0.5) unit-cell for the states with polarization pointing away from (solid) and toward (dashed) the interface. FM ordering is assumed throughout the supercell. The vertical dashed line denotes the Fermi energy. The labels indicate the character of the states which are either Mn d-states ($e_g$ or $t_{2g}$) or oxygen p-states (O-p).

In La$_{1-x}$A$_x$MnO$_3$ systems the majority Mn-$t_{2g}$ states are completely filled, whereas the minority $t_{2g}$ states are empty, contributing 3$\mu_B$ to the Mn magnetic moment. The hole concentration, $x$, determines the population of the majority-spin $e_g$ states, which lie above the $t_{2g}$ states because of the octahedral crystal field, making the overall local Mn magnetic moment formally $(4-x)\mu_B$. Examining the LDOS of the LAMO interface (Fig. 4) we see that the screening charge that accumulates plays a similar role as explicit doping in the bulk compounds. Indeed, Table I shows a significant change in the magnitude of the magnetic moments on the Mn sites as the polarization is reversed. For the ground state, the change in the magnitude of the Mn moment is 0.46 $\mu_B$ on the first Mn layer, 0.35 $\mu_B$ on the second layer, and 0.10 $\mu_B$ on the third layer.

Most importantly the $e_g$ states also determine the intersite magnetic interactions in the doped La-manganites, leading to the change in magnetic ground state orderings across the compositional phase diagram in the bulk materials.[35] To see that the magnetic reconstruction follows from the change in $e_g$ population, we make a comparison of our interface calculations with the magnetic phases of LAMO. The total energies of three types of collinear AFM order (A-, C- and G-type) of bulk LAMO are calculated and compared to the FM state. For simplicity, all the configurations we studied assume the same tetragonal structure that we use in the supercell.[55] The results are plotted in Fig. 5. Since we restrict our calculations to an undistorted perovskite structure we do not predict the A-type AFM order for $x \sim 0.0$, which is known to arise due to the sizable lattice distortions.[56] Here, however, we need to only focus on the region where there is a transition from FM metal to A-type AFM metal just above $x = 0.5$. This is consistent with previous DFT calculations of the doped La-manganites [50,51] and is in excellent agreement with the experimental phase diagram of $La_{1-x}Sr_xMnO_3$ where it is known this phase boundary persists up to about 240K.[40]

For polarization pointing into the interface we find that the $e_g$ population on a given layer is increased by, at most, ~0.23$e$. From Fig. 5 we expect this to enhance the stability of the FM state, which is in agreement with our calculations for the interface (see Table I). For polarization pointing away from the interface the population of the $e_g$ states decreased by ~0.23$e$, ~0.10$e$ and ~0.05$e$ on the first, second and third Mn layers, respectively. Examining Fig. 5 we therefore expect a transition to A-type order extending to the first 2 or 3 unit-cells, consistent with the results presented in Table I. Also from Fig. 5 we see that a reduction in $e_g$ population of at least ~0.3$e$ is required to induce at transition to C-type order. The change of ~0.23$e$ found in our supercell calculations is therefore consistent with the absence of a C-type transition in Table I. It is important to note, however, that the comparison between the bulk and surface is qualitative at best, especially for the Mn moments at the interface which have a local environment that is significantly different from the bulk. Indeed it was previously predicted that magnetic order differing from the bulk can arise spontaneously at manganite surfaces[57] and interfaces.[53]

It was previously suggested that polar distortions at the surfaces of the La-manganites lead to a substantial change in the intersite magnetic coupling, and therefore perhaps a change in the magnetic ordering.[49] To confirm that the change in magnetic order in our system is mainly due to the effects of electronic screening in the LAMO we perform an additional set of supercell calculations where the distortions in the LAMO (see Fig. 2) are suppressed, but keeping the ferroelectric displacements in the BTO. We find that the "frozen" $A_2$ magnetic configuration had energy –24 meV relative to the "frozen" FM configuration for polarization pointing away from the interface, whereas the "frozen" FM state is still stable for the polarization pointing into the interface. This confirms that the mechanism of electronic screening is the dominant contribution to the magnetic reconstruction.

Examining Fig. 5 we can also expect that for a stoichiometry further from the AF-FM transition that a magnetic reconstruction will not occur. To confirm this we performed calculations on a supercell with $x = 0.33$. We find that for both polarization orientations the $A_1$ configuration yields an increase in energy compared to the FM state. Revealingly, however, for polarization pointing away from the interface the $A_1$ configuration is only 1 meV above the FM state, whereas the opposite polarization state is 90 meV above. Because of this we expect that magnetic reconstruction should occur for slightly higher doping levels than $x = 0.33$. Therefore the effect is not limited to the $x = 0.5$ case, but rather is expected for a *range* of stoichiometries near the magnetic transition.

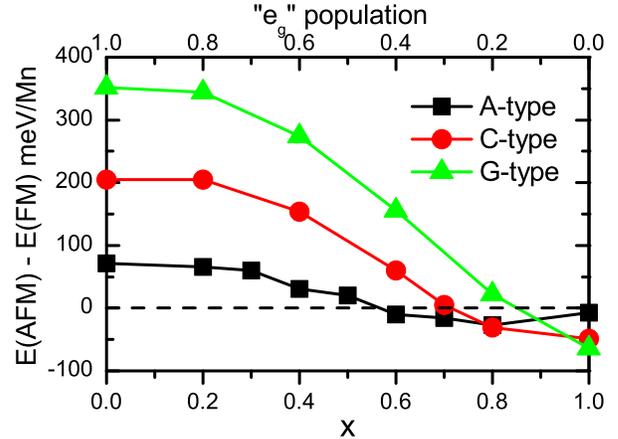

FIG. 5. Calculated energies of the various AFM configurations of bulk $La_{1-x}A_xMnO_3$ relative to the FM state as a function of $A$ concentration, $x$. The top scale gives the formal population of the majority channel Mn-$e_g$ states, $(1 - x)$.

## V. SUMMARY AND OUTLOOK

We have explored a different type of interface magnetoelectric effect at the interface between a metallic doped La-manganite and a ferroelectric oxide. We demonstrate this effect using first-principles calculations at the interface between $La_{1-x}A_xMnO_3$ and $BaTiO_3$, where $A$ is a divalent cation. By choosing $x$ near the magnetic phase transition we found that the magnetic configuration of the Mn moments near the interface changes from ferromagnetic ordering to antiferromagnetic ordering as the ferroelectric polarization is reversed. The origin of this effect is the modulation of the charge density induced on the interfacial LAMO layers to screen the polarization charges of the BTO. This magnetic reconstruction at the interface is entirely analogous to the magnetic phase transitions with change in composition of the doped La-manganites. Such a phenomenon constitutes a substantial and robust effect that is of considerable interest to the area of electrically controlled magnetism.

Finally, we would like to mention that the feasibility of observing the predicted phenomenon is corroborated by recent experiments on systems with ferroelectric $PbZr_{0.2}Ti_{0.8}O_3$ (PZT) in contact with a thin layer of either $La_{0.85}Ba_{0.15}MnO_3$[58] or $La_{0.8}Sr_{0.2}MnO_3$.[34] It was demonstrated that the magnetization as well as the ferromagnetic Curie temperature of the manganite change in response to reversal of the ferroelectric polarization. Both of these effects are consistent with modulating the charge around $x = 0.2$ of the phase diagram of the $La_{1-x}A_xMnO_3$ compounds, where it is not necessarily expected that a magnetic transition should occur. It would be interesting to perform experiments for magnetic oxide materials where the doping level is intentionally chosen to be close to the transition point between two different magnetic phases, as we have done here. An excellent choice in this regard would likely be $La_{1-x}Sr_xMnO_3$ with its FM metal to A-type AFM metal transition around $x = 0.5$.[40] Another interesting experimental possibility is to tune the stoichiometry only at the interface[59] to reside near the magnetic transition. This would make it possible to maintain a robust ferromagnetic metal throughout the bulk of the manganite, while optimizing the interface for the magnetic reconstruction effect. We therefore hope that the theoretical predictions described in this work will stimulate further experiments in this exciting field.

## ACKNOWLEDGEMENTS

This work was supported by the NSF-funded MRSEC (grant No. DMR-0820521), the Nanoelectronics Research Initiative of the Semiconductor Research Corporation and the Nebraska Research Initiative. Computations were performed utilizing the Research Computing Facility at UNL and the Center for Nanophase Materials Sciences at Oak Ridge National Laboratory.

† e-mail: jdburton1@gmail.com
‡ e-mail: tsymbal@unl.edu

[1] M. Fiebig, J. Phys. D, R123 (2005).
[2] W. Eerenstein, N. D. Mathur and J. F. Scott, Nature **442**, 759 (2006).
[3] H. Schmid, J. Phys.: Cond. Mat., 434201 (2008).
[4] M. Weisheit, S. Fahler, A. Marty, Y. Souche, C. Poinsignon and D. Givord, Science **315**, 349 (2007).
[5] D. Chiba, M. Sawicki, Y. Nishitani, Y. Nakatani, F. Matsukura and H. Ohno, Nature **455**, 515 (2008).
[6] C.-G. Duan, J. P. Velev, R. F. Sabirianov, W. N. Mei, S. S. Jaswal and E. Y. Tsymbal, Appl. Phys. Lett. **92**, 122905 (2008).
[7] T. Maruyama, Y. Shiota, T. Nozaki, K. Ohta, N. Toda, M. Mizuguchi, A. A. Tulapurkar, T. Shinjo, M. Shiraishi, S. Mizukami, Y. Ando and Y. Suzuki, Nature Nano. **4**, 158 (2009).
[8] P. Borisov, A. Hochstrat, X. Chen, W. Kleemann and C. Binek, Phys. Rev. Lett. **94**, 117203 (2005).
[9] V. Laukhin, V. Skumryev, X. Marti, D. Hrabovsky, F. Sanchez, M. V. Garcia-Cuenca, C. Ferrater, M. Varela, U. Luders, J. F. Bobo and J. Fontcuberta, Phys. Rev. Lett. **97**, 227201 (2006).
[10] M. Y. Zhuravlev, S. S. Jaswal, E. Y. Tsymbal and R. F. Sabirianov, Appl. Phys. Lett. **87**, 222114 (2005).
[11] E. Y. Tsymbal and H. Kohlstedt, Science **313**, 181 (2006).
[12] M. Gajek, M. Bibes, S. Fusil, K. Bouzehouane, J. Fontcuberta, A. Barthelemy and A. Fert, Nature Mater. **6**, 296 (2007).
[13] S. Ju, T.-Y. Cai, G.-Y. Guo and Z.-Y. Li, Phys. Rev. B **75**, 064419 (2007).
[14] J. P. Velev, C.-G. Duan, J. D. Burton, A. Smogunov, M. K. Niranjan, E. Tosatti, S. S. Jaswal and E. Y. Tsymbal, Nano Lett. **9**, 427 (2009).
[15] D. Khomskii, Physics **2**, 20 (2009).
[16] K. F. Wang, J. M. Liu and Z. F. Ren, Advances in Physics **58**, 321 (2009).
[17] L. D. Landau and E. M. Lifshitz, *Classical Theory of Fields* (Pergamon, Oxford, 1975).
[18] I. E. Dzyaloshinskii, Sov. Phys. JETP **10**, 628 (1960).
[19] C. J. Fennie and K. M. Rabe, Phys. Rev. Lett. **97**, 267602 (2006).
[20] H. Zheng, J. Wang, S. E. Lofland, Z. Ma, L. Mohaddes-Ardabili, T. Zhao, L. Salamanca-Riba, S. R. Shinde, S. B. Ogale, F. Bai, D. Viehland, Y. Jia, D. G. Schlom, M. Wuttig, A. Roytburd and R. Ramesh, Science **303**, 661 (2004).
[21] R. V. Chopdekar and Y. Suzuki, Appl. Phys. Lett. **89**, 182506 (2006).
[22] W. Eerenstein, M. Wiora, J. L. Prieto, J. F. Scott and N. D. Mathur, Nature Mater. **6**, 348 (2007).
[23] S. Sahoo, S. Polisetty, C.-G. Duan, S. S. Jaswal, E. Y. Tsymbal and C. Binek, Phys. Rev. B **76**, 092108 (2007).
[24] R. Ramesh and N. A. Spaldin, Nature Mater. **6**, 21 (2007).
[25] C.-G. Duan, S. S. Jaswal and E. Y. Tsymbal, Phys. Rev. Lett. **97**, 047201 (2006).
[26] M. Fechner, I. V. Maznichenko, S. Ostanin, A. Ernst, J. Henk, P. Bruno and I. Mertig, Phys. Rev. B **78**, 212406 (2008).
[27] K. Yamauchi, B. Sanyal and S. Picozzi, Appl. Phys. Lett. **91**, 062506 (2007).
[28] M. K. Niranjan, J. P. Velev, C.-G. Duan, S. S. Jaswal and E. Y. Tsymbal, Phys. Rev. B **78**, 104405 (2008).
[29] S. Zhang, Phys. Rev. Lett. **83**, 640 (1999).
[30] C.-G. Duan, J. P. Velev, R. F. Sabirianov, Z. Zhu, J. Chu, S. S. Jaswal and E. Y. Tsymbal, Phys. Rev. Lett. **101**, 137201 (2008).
[31] J. M. Rondinelli, M. Stengel and N. A. Spaldin, Nature Nano. **3**, 46 (2008).
[32] C.-G. Duan, C.-W. Nan, S. S. Jaswal and E. Y. Tsymbal, Phys. Rev. B **79**, 140403 (2009).
[33] M. K. Niranjan, J. D. Burton, J. P. Velev, S. S. Jaswal and E. Y. Tsymbal, Appl. Phys. Lett. **95**, 052501 (2009).
[34] H. J. A. Molegraaf, J. Hoffman, C. A. F. Vaz, S. Gariglio, D. v. d. Marel, C. H. Ahn and J.-M. Triscone, Advanced Materials **21**, 3470 (2009).
[35] E. Dagotto, T. Hotta and A. Moreo, Phys. Rep. **344**, 1 (2001).
[36] P. G. de Gennes, Phys. Rev. **118**, 141 (1960).
[37] C. H. Ahn, J. M. Triscone and J. Mannhart, Nature **424**, 1015 (2003).
[38] A supercell with 9.5 unit cells of LAMO displayed nearly identical lattice relaxations and magnetic energies for the A-type configurations as for the supercell with 5.5 formula units of LAMO, but was too large for computations of the C- and G-type magnetic configurations. In particular, the magnetic energies in Table I differed by less than 1meV.
[39] L. Nordheim, Annalen der Physik **401**, 607 (1931).
[40] T. Akimoto, Y. Maruyama, Y. Moritomo, A. Nakamura, K. Hirota, K. Ohoyama and M. Ohashi, Phys. Rev. B **57**, R5594 (1998).
[41] P. Giannozzi, S. Baroni, N. Bonini, M. Calandra, R. Car, C. Cavazzoni, D. Ceresoli, G. L. Chiarotti, M. Cococcioni, I. Dabo, A. D. Corso, S. d. Gironcoli, S. Fabris, G. Fratesi, R. Gebauer, U. Gerstmann, C. Gougoussis, A. Kokalj, M. Lazzeri, L. Martin-Samos, N. Marzari, F. Mauri, R. Mazzarello, S. Paolini, A. Pasquarello, L. Paulatto, C. Sbraccia, S. Scandolo, G. Sclauzero, A. P. Seitsonen, A. Smogunov, P. Umari and R. M. Wentzcovitch, J. Phys.: Cond. Mat. **21**, 395502 (2009).
[42] J. P. Perdew, K. Burke and M. Ernzerhof, Phys. Rev. Lett. **77**, 3865 (1996).

[43] D. Vanderbilt, Phys. Rev. B **41**, 7892 (1990).

[44] D. Vanderbilt, http://www.physics.rutgers.edu/~dhv/uspp/.

[45] In the supercell we do not account for the rotational distortions of the octahedral oxygen cages around the Mn-ions in the LAMO. We find these distortions do not affect appreciably the stability of the bulk magnetic phases for $x = 0.3 - 0.7$, which is the relevant range of x discussed here.

[46] J. Zhang, H. Tanaka, T. Kanki, J.-H. Choi and T. Kawai, Phys. Rev. B **64**, 184404 (2001).

[47] A. Ciucivara, B. Sahu and L. Kleinman, Phys. Rev. B **77**, 092407 (2008).

[48] K. J. Choi, M. Biegalski, Y. L. Li, A. Sharan, J. Schubert, R. Uecker, P. Reiche, Y. B. Chen, X. Q. Pan, V. Gopalan, L. Q. Chen, D. G. Schlom and C. B. Eom, Science **306**, 1005 (2004).

[49] J. M. Pruneda, V. Ferrari, R. Rurali, P. B. Littlewood, N. A. Spaldin and E. Artacho, Phys. Rev. Lett. **99**, 226101 (2007).

[50] W. Luo, A. Franceschetti, M. Varela, J. Tao, S. J. Pennycook and S. T. Pantelides, Phys. Rev. Lett. **99**, 036402 (2007).

[51] Y. Konishi, Z. Fang, M. Izumi, T. Manako, M. Kasai, H. Kuwahara, M. Kawasaki, K. Terakura and Y. Tokura, J. Phys. Soc. Jpn. **68**, 3790 (1999).

[52] A. Filippetti and W. E. Pickett, Phys. Rev. B **62**, 11571 (2000).

[53] W. Luo, S. J. Pennycook and S. T. Pantelides, Phys. Rev. Lett. **101**, 247204 (2008).

[54] T. Hotta, S. Yunoki, M. Mayr and E. Dagotto, Phys. Rev. B **60**, R15009 (1999).

[55] Total energy calculations of the FM, A-type, C-type and G-type states were performed using 20x20x20, 20x20x10, 14x14x20 and 14x14x10 k-point grids, respectively.

[56] H. Sawada, Y. Morikawa, K. Terakura and N. Hamada, Phys. Rev. B **56**, 12154 (1997).

[57] A. Filippetti and W. E. Pickett, Phys. Rev. Lett. **83**, 4184 (1999).

[58] T. Kanki, H. Tanaka and T. Kawai, Appl. Phys. Lett. **89**, 242506 (2006).

[59] Y. Hikita, M. Nishikawa, T. Yajima and H. Y. Hwang, Phys. Rev. B **79**, 073101 (2009).